# A Fast Affine Projection Algorithm Based on Matching Pursuit in Adaptive Noise Cancellation for Speech Enhancement

*N. Sonbolestan*
School of Electrical & Electronic Engineering
The University of Manchester
Manchester, United Kingdom
e-mai-
Noushin.Sonbolestan@postgrad.manchester.ac.uk

*S. A. Hadei*
Department of Electrical Engineering
Tarbiat Modares University
Tehran, Iran
e-mail- a.hadei@modares.ac.ir

*Abstract*— In many application of noise cancellation, the changes in signal characteristics could be quite fast. This requires the utilization of adaptive algorithms, which converge rapidly. Least Mean Squares (LMS) adaptive filters have been used in a wide range of signal processing application. The Recursive Least Squares (RLS) algorithm has established itself as the "ultimate" adaptive filtering algorithm in the sense that it is the adaptive filter exhibiting the best convergence behavior. Unfortunately, practical implementations of the algorithm are often associated with high computational complexity and/or poor numerical properties. Recently adaptive filtering was presented that was based on Matching Pursuits, have a nice tradeoff between complexity and the convergence speed. This paper describes a new approach for noise cancellation in speech enhancement using the new adaptive filtering algorithm named fast affine projection algorithm (FAPA). The simulation results demonstrate the good performance of the FAPA in attenuating the noise.

## I. INTRODUCTION

It is well known that two of most frequently applied algorithms for noise cancellation [1] are normalized least mean squares (NLMS) [2]-[5] and recursive least squares (RLS) [6]-[10] algorithms. Considering these two algorithms, it is obvious that NLMS algorithm has the advantage of low computational complexity. On the contrary, the high computational complexity is the weakest point of RLS algorithm but it provides a fast adaptation rate. Thus, it is clear that the choice of the adaptive algorithm to be applied is always a tradeoff between computational complexity and fast convergence. The adaptive filter algorithm based on the matching pursuit (MP) was presented in [11]-[13]. This algorithm which was called Fast Affine Projection (FAP) algorithm was fully developed in [18]. The convergence property of the FAP algorithm is superior to that of the usual LMS, NLMS algorithms and comparable to that of the RLS algorithm [14]. In this algorithm, one of the filter coefficients is updated one or more at each time instant, in order to fulfill a suitable tradeoff between convergences rate and computational complexity [18]. The performance of the proposed algorithm is fully studied through the energy conservation [15], [16] analysis used in adaptive filters and the general expressions [17] for the steady-state mean square error and transient performance analysis were derived in [18].

What we propose in this paper is the use of the FAP algorithm in noise cancellation for speech enhancement. We compare the results with classical adaptive filter algorithm such as LMS, NLMS, and RLS algorithms. Simulation results show the good performance of the FAP algorithm in attenuating the noise. In the following we find also the optimum parameter which is used in this algorithm.

We have organized our paper as follows:

In the next section, the classical adaptive algorithms such as LMS, NLMS and RLS algorithms will be reviewed. In the following the FAP algorithm in [18] will be briefly introduced. Section 4 presents the adaptive noise cancellation setup. We conclude the paper with comprehensive set of simulation results.

Throughout the paper, the following notations are adopted:

TABLE I. TABLE TYPE STYLES

| | |
|---|---|
| $\|\cdot\|$ | Norm of a scalar |
| $\|\cdot\|^2$ | Squared Euclidean norm of a vector |
| $(.)^T$ | Transpose of a vector or a matrix |
| $(.)^{-1}$ | Inverse of a scalar or a matrix |
| $<.,.>$ | Inner product of two vectors. |

## II. BACKGROUND ON LMS, NLMS AND RLS ALGORITHM

Fig. 1, we show the prototypical adaptive filter setup, where $x(n)$, $d(n)$ and $e(n)$ are the input, the desired and the output error signals, respectively. The vector $\underline{h}(n)$ is the $M \times 1$ column vector of filter coefficient at time $n$, in such away that the output of signal, $y(n)$, is good estimate of the desired signal, $d(n)$.

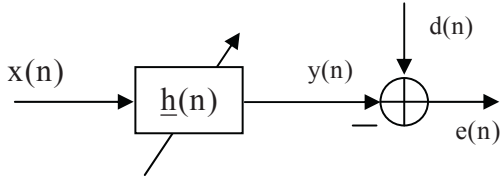

Fig.1. Prototypical adaptive filter setup

It is well known that the filter vector update equation for the LMS algorithm is given by [9]:

$$\underline{h}(n+1) = \underline{h}(n) + \mu \underline{x}(n)e(n), \quad (1)$$

where

$$\underline{x}(n) = [x(n), x(n-1), \ldots, x(n-M+1)]^T, \quad (2)$$

and $\mu$ is the step-size that determines the convergence speed and steady-state mean-square error (MSE). Also, the output error signal, $e(n)$, is given by

$$e(n) = d(n) - \underline{h}^T(n)\underline{x}(n). \quad (3)$$

To increase the convergence speed of the LMS algorithm, the NLMS algorithm was proposed which can be stated as [9]

$$\underline{h}(n+1) = \underline{h}(n) + \frac{\mu}{\|\underline{x}(n)\|^2}\underline{x}(n)e(n) \quad (4)$$

The filter vector update equation in RLS algorithm is [14]:

$$\underline{h}(n+1) = \underline{h}(n) + C^{-1}(n)\underline{x}(n)e(n), \quad (5)$$

where $C(n)$ is the estimation of the autocorrelation matrix. This matrix is given by

$$C(n) = \sum_{i=0}^{n} \lambda^{n-i} \underline{x}(i)\underline{x}^T(i). \quad (6)$$

The $\lambda$ parameter is the forgetting factor and $0 \ll \lambda < 1$.

## III. AFFINE PROJECTION ALGORITHM (FAPA)

### A. Notation and problem description

With reference to Figure 1, the error signal, $e(n)$, can be expressed as:

$$e(n) = d(n) - \sum_{k=0}^{M-1} h_k(n)x(n-k). \quad (7)$$

Considering the samples $n-L+1, n-L+2, \ldots, n$, where we focus on the situation where $L > M$, Eq.7 can be written as:

$$\underline{e}(n) = \underline{d}(n) - X(n)\underline{h}(n), \quad (8)$$

where

$$X(n) = [\underline{x}_0(n), \underline{x}_1(n), \ldots, \underline{x}_{M-1}(n)]. \quad (9)$$

These columns are furthermore defined through

$$\underline{x}_j(n) = [x(n-j), x(n-j-1), \ldots, x(n-j-L+1)]^T. \quad (10)$$

The vector of desired signal samples is given by

$$\underline{d}(n) = [d(n), d(n-1), \ldots, d(n-L+1)]^T, \quad (11)$$

and $\underline{e}(n)$ is defined similarly. The adaptive filtering problem can now be formulated as the task of finding the update for $\underline{h}(n)$, at each time instant $n$, such that the error is made as small as possible.

Note that $X(n)\underline{h}(n)$ can be written as

$$X(n)\underline{h}(n) = \sum_{k=0}^{M-1} h_k(n)\underline{x}_k(n), \quad (12)$$

i.e. as a weighted sum of the columns of $X(n)$ with the elements of $\underline{h}(n)$ being the weighting factors. A greedy algorithm for successively building (better) approximations to a given vector using linear combinations of vectors from a given set is the BMP algorithm [19]. Inspired by this algorithm, conceived and developed in another context and with other motivations than those of this paper, we devise a procedure for recursively building an approximation to $\underline{d}(n)$ using linear combinations of the columns of $X(n)$.

### B. Algorithm development

Assuming that we have an approximation to $\underline{d}(n-1)$ at time $n-1$ given by $X(n-1)\underline{h}(n-1)$, the *apriori* approximation error at time $n$ is

$$\underline{e}_o(n) = \underline{d}(n) - X(n)\underline{h}(n-1). \quad (13)$$

In building a better approximation through the update of only one coefficient in $\underline{h}(n-1)$, we would write the new error as

$$\underline{e}_1(n) = \underline{d}(n) - (X(n)\underline{h}(n-1) + X(n)h_{j_o(n)}^{update}(n)\underline{u}_{j_o(n)}) \quad (14)$$

Note that $j_o(n)$ is the index of the coefficient to be update in the zero'th P-iteration at time $n$, and $\underline{u}_j$ is the M-vector with 1 in position $j$ and 0 in all other positions. Intuitively, it would make sense to select $j_o(n)$ as the index corresponding to that column of $X(n)$ that is most similar to the apriori approximation error of Eq. 13. Thus, $j_o(n)$ is found as the index of the column of $X(n)$ onto which $\underline{e}_o(n)$ has its maximum projection, -or in other words:

$$j_o(n) = \arg\max_j \frac{|\langle \underline{e}_o(n), \underline{x}_j(n)\rangle|}{\|\underline{x}_j(n)\|}, \quad (15)$$

Where $<.,.>$ denotes an inner product between the two vector arguments. Given the index $j_o(n)$, the update of the corresponding filter coefficient is

$$h_{j_o(n)}(n) = h_{j_o(n)}(n-1) + h^{update}_{j_o(n)}(n), \quad (16)$$

where $h^{update}_{j_o(n)}(n)$ is the value of the projection of $\underline{e}_o(n)$ onto the unit vector with direction given by $\underline{x}_{j_o(n)}(n)$, i.e.:

$$h^{update}_{j_o(n)}(n) = \frac{<\underline{e}_o(n), \underline{x}_{j_o(n)}(n)>}{\left\|\underline{x}_{j_o(n)}(n)\right\|^2}. \quad (17)$$

Thus, the zero'th P-iteration updates the filter vector as follows:

$$\underline{h}^{(o)}(n) = \underline{h}(n-1) + h^{update}_{j_o(n)}(n)\underline{u}_{j_o(n)}. \quad (18)$$

To have control on the convergence speed and stability of the algorithms, we introduce the step-size in the algorithm as following:

$$\underline{h}^{(o)}(n) = \underline{h}(n-1) + \mu h^{update}_{j_o(n)}(n)\underline{u}_{j_o(n)} \quad (19)$$

Given this, the updated error expression of Eq.14 can be written as:

$$\underline{e}_1(n) = \underline{d}(n) - X(n)\underline{h}^{(o)}(n). \quad (20)$$

If we want to do more than one P-iteration at time $n$, the procedure described above starting with finding the maximum projection of $\underline{e}_o(n)$ onto a column of $X(n)$ can be repeated with $\underline{e}_1(n)$ taking the role of $\underline{e}_o(n)$. This can be repeated as many times as desired, say P times, leading to a sequence of coefficient updates:

$$h_{j_o(n)}(n), h_{j_1(n)}(n), \ldots, h_{j_{P-1}(n)}(n). \quad (21)$$

Note that if $P > 2$ it is entirely possible that one particular coefficient is updated more than once at a given time $n$. The resulting filter coefficient vector after P iterations at time $n$ is denoted $\underline{h}^{(P-1)}(n)$, but where there is no risk of ambiguity, we shall refer to this filter vector simply as $\underline{h}(n)$.

The procedure described above corresponds to applying the BMP algorithm [19] to a dictionary of vectors given by the columns of $X(n)$ for the purpose of building an approximation to $\underline{d}(n)$. The only difference is that we do this for each new time instant $n$ while keeping the results of the BMP from the previous time instant $n-1$. It is interesting to note that a slightly different, but equivalent, procedure to the one described above would result if we tried to find the least squares solution to the over determined set of equations (remember $L > M$):

$$X(n)\underline{h}(n) = \underline{d}(n) \quad (22)$$

Subject to the constrain that, given an initial solution, say $\underline{h}_o(n)$, we are allowed to adjust only one element of this vector.

From the above, it is evident that the key computations of our adaptive filter algorithm are those of Eqs.15 and 17. Making use of Eqs. 13 and 12, we find

$$j_o(n) = \arg\max_j \frac{1}{\left\|\underline{x}_j(n)\right\|} \Big| <\underline{d}(n), \underline{x}_j(n)> - \sum_{k=o}^{M-1} h_k(n-1)<\underline{x}_k(n), \underline{x}_j(n)> \Big| \quad (23)$$

and

$$h^{update}_{j_o(n)}(n) = \frac{1}{\left\|\underline{x}_{j_o(n)}(n)\right\|^2}\{<\underline{d}(n), \underline{x}_{j_o(n)}(n)> - \sum_{k=o}^{M-1} h_k(n-1)<\underline{x}_k(n), \underline{x}_{j_o(n)}(n)>\}. \quad (24)$$

These are the pertinent equations if one coefficient update, i.e. one P-iteration is performed for each new signal sample. Note that having computed the terms of Eq. 23, very little additional work is involved in finding the update of Eq. 24. It is instructive to explicitly state these equations also for iteration no. $i > 0$ at time $n$:

$$j_i(n) = \arg\max_j \frac{1}{\left\|\underline{x}_j(n)\right\|} \Big| <\underline{d}(n), \underline{x}_j(n)> - \sum_{k=o}^{M-1} h_k^{(i-1)}(n)<\underline{x}_k(n), \underline{x}_j(n)> \Big| \quad (25)$$

and

$$h^{update}_{j_i(n)}(n) = \frac{1}{\left\|\underline{x}_{j_i(n)}(n)\right\|^2}\{<\underline{d}(n), \underline{x}_{j_i(n)}(n)> - \sum_{k=o}^{M-1} h_k^{(i-1)}(n)<\underline{x}_k(n), \underline{x}_{j_i(n)}(n)>\}. \quad (26)$$

From these equations it is evident that some terms depend only on $n$, i.e. they need to be computed once for each $n$ and can subsequently be used unchanged for all P-iterations at time $n$. Other terms depend on both *n* and the P-iteration index and must consequently be updated for each P-iteration. Since we must associate the update depending only on n with iteration no. 0, this is the computationally most expensive update.

From the above it is evident that the inner products

$<\underline{d}(n), \underline{x}_j(n)>$ and $<\underline{x}_k(n), \underline{x}_j(n)>$ play prominent roles in the computations involved in the algorithm. As formulated up to this point, obvious recursions for these inner products are

$$<\underline{d}(n), \underline{x}_j(n)> = <\underline{d}(n-1), \underline{x}_j(n-1)> + d(n)x(n-j) - d(n-L)x(n-j-L) \quad (27)$$

and

$$<\underline{x}_k(n), \underline{x}_j(n)> = <\underline{x}_k(n-1), \underline{x}_j(n-1)> + x(n-k)x(n-j) - x(n-k-L)x(n-j-L) \quad (28)$$

## IV. ADAPTIVE NOISE CANCELLATION

Fig. 2 shows the adaptive noise cancellation setup. In this application, the corrupted signal passes through a filter that tends to suppress the noise while leaving the signal unchanged. This process is an adaptive process, which means it cannot require a priori knowledge of signal or noise characteristics. Adaptive noise cancellation algorithms utilize two or more microphones (sensor). One microphone is used to measure the speech + noise signal while the other is used to measure the noise signal alone. The technique adaptively adjusts a set of filter coefficients so as to remove the noise from the noisy signal. This technique, however, requires that the noise component in the corrupted signal and the noise in the reference channel have high coherence. Unfortunately this is a limiting factor, as the microphones need to be separated in order to prevent the speech being included in the noise reference and thus being removed. With large separations the coherence of the noise is limited and this limits the effectiveness of this technique. In summary, to realize the adaptive noise cancellation, we use two inputs and an adaptive filter. One input is the signal corrupted by noise (Primary Input, which can be expressed as $s(n) + n_0(n)$). The other input contains noise related in some way to that in the main input but does not contain anything related to the signal (Noise Reference Input, expressed as $n_1(n)$). The noise reference input pass through the adaptive filter and output $y(n)$ is produced as close a replica as possible of $n_0(n)$. The filter readjusts itself continuously to minimize the error between $n_0(n)$ and $y(n)$ during this process. Then the output $y(n)$ is subtracted from the primary input to produce the system output $e = s + n_0 - y$, which is the denoised signal. Assume that $s$, $n_0$, $n_1$ and $y$ are statistically stationary and have zero means. Suppose that $s$ is uncorrelated with $n_0$ and $n_1$, but $n_1$ is correlated with $n_0$. We can get the following equation of expectations:

$$E[e^2] = E[s^2] + E[(n_0 - y)^2] \quad (29)$$

When the filter is adjusted so that $E[e^2]$ is minimized, $E[(n_0 - y)^2]$ is also minimized. So the system output can serve as the error signal for the adaptive filter. The adaptive noise cancellation configuration is shown in Fig. 2. In this setup, we model the signal path from the noise source to primary sensor as an unknown FIR channel $W_e$. Applying the adaptive filter to reference noise at reference sensor, we then employ an adaptive algorithm to train the adaptive filter to match or estimate the characteristics of unknown channel $W_e$.

If the estimated characteristics of unknown channel have negligible differences compared to the actual characteristics, we should be able to successfully cancel out the noise component in corrupted signal to obtain the desired signal. Notice that both of the noise signals for this configuration need to be uncorrelated to the signal $s(n)$. In addition, the noise sources must be correlated to each other in some way, preferably equal, to get the best results.

Do to the nature of the error signal, the error signal will never become zero. The error signal should converge to the signal $s(n)$, but not converge to the exact signal. In other words, the difference between the signal $s(n)$ and the error signal $e(n)$ will always be greater than zero. The only option is to minimize the difference between those two signals.

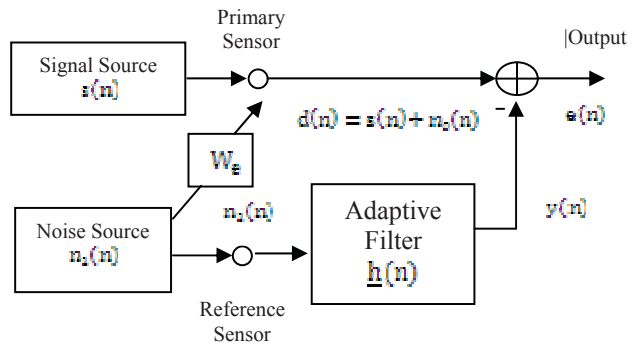

Fig. 2. Adaptive Noise Cancellation Setup

## V. EXPERIMENTAL RESULTS

In this section we evaluate the performance of each algorithm in noise cancellation setup as shown in Fig. 2. The original, primary, and reference signals are from the reference [20]. The original speech is corrupted with office noise. The signal to noise ratio (SNR) of the primary signal is -10.2180dB. This signal is then processed as in Fig. 2. Fig. 3 shows the signals.

The order of the filter was set to M=8. The parameter $\mu$ was set to 0.002 in the LMS and 0.005 in the NLMS algorithm. Fig. 4 shows the filtered output signal and the mean squared error (learning curve) in the LMS algorithm. The SNR of the filtered signal is calculated for this experiment. The SNR improvement (SNRI) is defined as the final SNR minus the original SNR. The SNRI in the LMS algorithm is 13.5474. Fig. 5 shows the results for NLMS algorithm. As we can see the convergence speed in the NLMS algorithm is faster than LMS algorithm. This fact can be seen in both filtered output and learning curve. For the NLMS algorithm the SNRI is 17.1056.

Fig. 6 shows the results for RLS algorithm. In this algorithm, the parameter $\lambda$ was set to 0.99. The results show that the RLS algorithm has faster convergence speed compared with LMS and NLMS algorithms. The SNRI in this algorithm is 29.7355.

In Figs. 7, we presented the results for FAP algorithms. The parameters was set to $L = 25, P = 8, \mu = 0.002$. The results show that the FAP has faster convergence speed than LMS and NLMS algorithms and comparable with the RLS algorithm. The SNRI in these algorithms is 21.0757. Table 2 summarizes the SNRI results.

TABLE II. SNR IMPROVEMENT IN DB

| Algorithm | LMS | NLMS | RLS | FAPA |
|---|---|---|---|---|
| SNRI(dB) | 13.5474 | 17.1056 | 29.7355 | 24.0757 |

## VI. CONCLUTIONS

In this paper we have applied a FAP algorithm on adaptive noise cancellation setup. The simulation results were compared with the classical adaptive filters, such as LMS, NLMS, and RLS algorithms, for attenuating noise in speech signals. In each algorithm the mean square error and the output of filter were presented. The simulation results show that the convergence rate of this algorithm is comparable with the RLS algorithm. Also, the optimum values of the FAP algorithm were calculated through experiments. In this algorithm, the number of iterations to be performed at each new sample time is a user selected parameter giving rise to an attractive and explicit tradeoff between convergence/tracking properties and computational complexity.

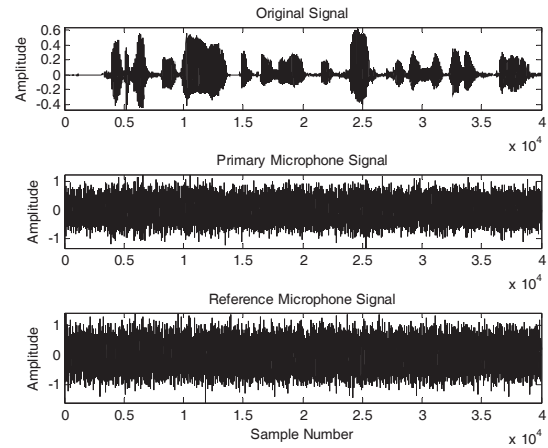

Fig. 3. Original, primary and reference signals.

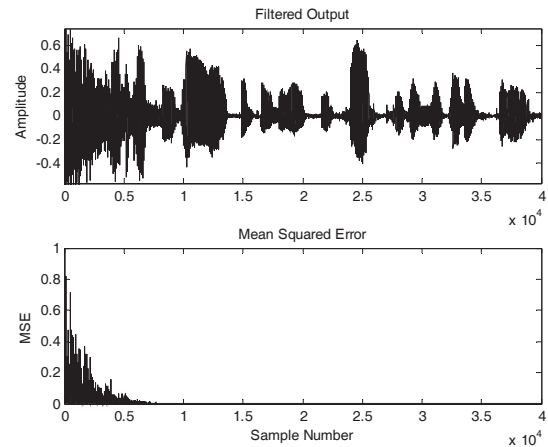

Fig. 4. Filtered output signal and MSE curve of the LMS algorithm.

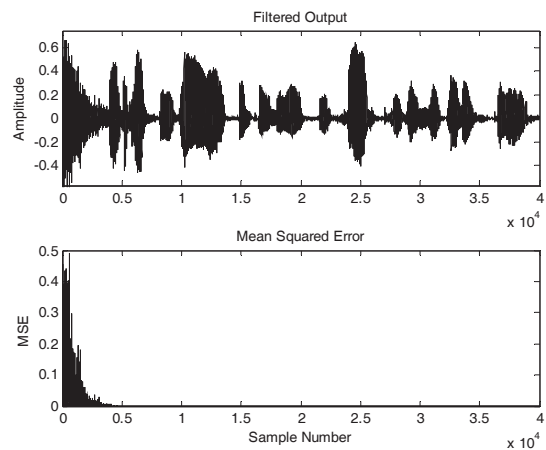

Fig. 5. Filtered output signal and MSE curve of the NLMS algorithm.

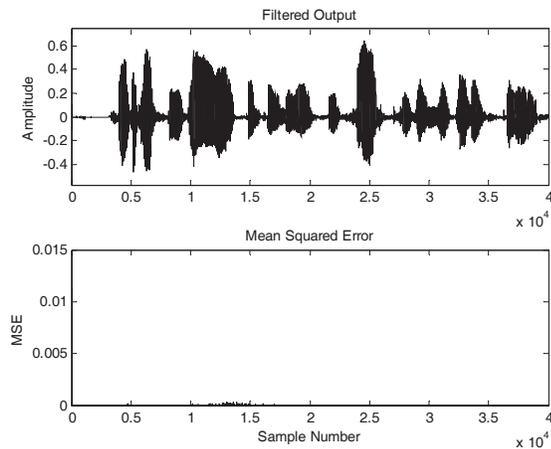

Fig. 6. Filtered output signal and MSE curve of the RLS algorithm.

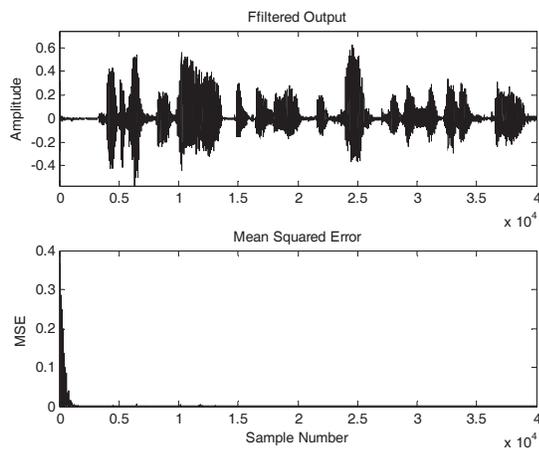

Fig. 7. Filtered output signal and MSE curve of the FAP algorithm.